\documentclass[twocolumn,twoside]{article}

\usepackage{graphicx}
 \usepackage{hyperref}

\begin{document}

\thispagestyle{plain}
\markboth{\rm \uppercase{Particle creation in the early
Universe}}{\rm \uppercase{Grib, Pavlov}}

\twocolumn[
\begin{center}
{\LARGE \bf Particle creation in the early Universe:\\[5pt]
achievements and problems}
\vspace{12pt}

{\large \bf {A. A. Grib${}^{1,2,\dag}$, \ Yu. V. Pavlov${}^{1,3,\ddag}$}}
\vspace{12pt}

{${}^{1}$\it A.\,Friedmann Laboratory for Theoretical Physics,
St.\,Petersburg, Russia;}
\vspace{4pt}

{\it ${}^{2}$Theoretical Physics and Astronomy Department,
The Herzen University,\\
Moika 48, St. Petersburg 191186, Russia;}
\vspace{4pt}

{\it ${}^{3}$Institute of Problems in Mechanical Engineering,
Russian Acad. Sci.,\\
Bol'shoy pr. 61, St. Petersburg 199178, Russia}
\end{center}
\vspace{5pt}
 {\bf Abstract.}
    Results on particle creation from vacuum by the gravitational field of
the expanding Friedmann Universe are presented.
    Finite results for the density of particles and the energy density for
created particles are given for different exact solutions for different
regimes of the expansion of the Universe.
    The  results are obtained as for conformal as for nonconformal particles.
    The hypothesis of the origination of visible matter from the decay of
created from vacuum superheavy particles identified with the dark matter is
discussed.

\vspace{11pt}
\vspace{17pt}
]

{\centering  \section{Introduction}}

\footnotetext[2]{andrei\_grib@mail.ru}
\footnotetext[3]{yuri.pavlov@mail.ru}

    The problem of particle creation from vacuum by the gravitational field
of the expanding Universe was studied intensively in the 70-ies of the past
century (see our works~\cite{GribMamayev69,GMM80} and papers of
Ya.B.\,Zel'dovich and
A.A.\,Starobinsky~\cite{ZeldovichStarobinsky71,Zeldovich}).
    High interest to the problem was expressed by
K.P.\,Staniukovich~\cite{StanukovichMelnikov}, who always supported one
of the authors of this paper (A.A.G.) in his activity.
    Really, differently from the usual opinion that speaking about the Big Bang
one must consider very high densities of matter composed from massive
particles  particle creation from vacuum at the certain epoch of the evolution
of the Universe makes possible to say that early Universe did not contain
massive particles and looked as vacuum and possibly some fields.
    In works~\cite{GribMamayev69,GMM80,MMSt} finite values for the density of
created particles and the energy density were obtained for the first time.
    It became clear that in the Universe described by the Friedmann metric
some special era exists for every type of particles defined by
the Compton time of the particle so that there is no creation of these
particles for the time much smaller or much larger than this time passed
from the beginning of the Universe.
    In papers~\cite{ZeldovichStarobinsky71,LukashStarobinsky74} an important
result was obtained that in the anisotropic Universe particle creation leads
to the isotropization of the metric which can be explanation of the fact
that our Universe is isotropic.

    However after these first successes the interest to the problem somehow
failed.
    The reason for this is the incorrect evaluation of the quantitative
meaning of the effect in  books of high
authority~\cite{ZeldovichNovikovSEU,WeinbergFfm} using the hypothesis of
the creation of only those particles which are observed today.
    Due to the fact that the effect is proportional to the mass of the
particle the effect is surely negligible for small masses.
    However if one supposes that observable particles appeared after
creation by the gravitation of massive particles with the mass of the order
of the Grand Unification the result is not small at all and even more it
makes possible to explain the observable number of particles in
the Universe~\cite{GMM,GD}.
    The other reason of the neglect is the incorrect statement in much
popular book~\cite{BD} that calculations based on the method of
the diagonization of the Hamiltonian lead to the infinite value of
the density of the created particles for nonconformal particles.
    This opinion was shown to be wrong by calculations in~\cite{Pavlov2001}.
    Note that in the book~\cite{BD} as well as in the book~\cite{WaldQFT}
there are no any finite (!) values of physical evaluations of the effect
of particle creation in cosmology.
    This occurred because of the fact that the method of the diagonalization
of the Hamiltonian strongly depends on the  form of the Hamiltonian used
for calculation of the particle creation in Friedmann Universe.
    The authors of~\cite{GribMamayev69} were lucky in 1969 to use the correct
expression of this Hamiltonian the ground state of which leads to the finite
value of the observable quantities.
    This Hamiltonian is defined by the metrical energy-momentum tensor of
the quantized massive field and is connected with the canonical Hamiltonian
of particles with varied mass in conformally connected to the Friedmann
metric static space.
    It is this connection which is leading to to finite results as for
conformal as for nonconformal particles.
    Any other choice of vacuum or Hamiltonian leads to the infinite value
of the density of created particles making the theory physically meaningless.
    In this paper after a short review of the results which can be found in
our book~\cite{GMM} we give a review of some new results obtained by
the authors in last time.
    Some unsolved problems are discussed in the Conclusion.

    We use the system of units in which $\hbar = c \!=\! 1$.\

\vspace{1em}
\section{Scalar field  in curved Friedmann space-time}
\hspace{\parindent}
     Consider the complex scalar field $\varphi(x)$ of the mass $m$
in curved Friedmann space-time with the Lagrangian
    \begin{equation}
L(x)=\sqrt{|g|} \left[\, g^{ik}\partial_i\varphi^*\partial_k\varphi -
(m^2 + \xi R)\, \varphi^* \varphi \right],
\label{Lag}
\end{equation}
    leading to Euler equations
\begin{equation}
( \nabla^i \nabla_{i} + \xi R + m^2 )\, \varphi(x)=0 ,
\label{Eqm}
\end{equation}
    where ${\nabla}_{\! i}$ are covariant derivatives in metric $g_{ik}$\
of the $N$-dimensional space-time,
$ g = {\rm det}(g_{ik})$,\ $R$ --- the curvature scalar.
    For $m=0$ and $\xi=\xi_c \equiv (N-2)/\,[ 4 (N-1)] $
equation~(\ref{Eqm}) is conformal invariant.
    This means that in mapping of the space-time with metric $g_{ik}$
on the space-time with metric ${\tilde g}_{ik}$:
    \begin{equation}
g_{ik} \to {\tilde g}_{ik} = \exp [-2\sigma(x)] \, g_{ik} ,
\label{cgik}
\end{equation}
    where $\sigma(x)$ is arbitrary smooth function of coordinates there
exists such scale transformation
    \begin{equation}
\varphi(x) \to {\tilde \varphi}(x) =
\exp \left[ \frac{N-2}{2}\, \sigma(x) \right] \varphi(x)  ,
\label{phitilde}
\end{equation}
    that the wave equation conserves its form (see~\cite{GMM}).
    The physical sense of the conformal invariance is that massless field
has no its proper length scale (for the massive case such a scale is given
by the Compton wave length $\lambda_C=1/m$) and so it must behave equally
in conformally connected spaces~(\ref{cgik}).
    In case $\xi=\xi_c $ equation~(\ref{Eqm}) is called equation with
conformal coupling ($\xi_c=1/6$ for $N=4$).
    The case $\xi=0$ corresponds to minimal coupling.
    Such type of  coupling  of the scalar field with curvature usually is
supposed in inflation models~\cite{Linde}.
    In quantum theory of fields in curved space-time one often considers
the case of arbitrary $\xi$~\cite{BD} (more general case of coupling with
the curvature of Gauss-Bonnet type was considered in~\cite{Pv4,Pavlov2013}).
    Then conformal invariance for massless fields is absent.
    The study of  nonconformal case is important due to different reasons.
    Massive vector mesons~\cite{GMM} (longitudinal components)
and gravitons~\cite{GrishchukY80} satisfy equations of this type.
    In case of the scalar field with self interaction in general it is
impossible to conserve conformal invariance not only of the effective action
(the conformal anomaly) but of the action itself~\cite{BD}.

    Note that the Dirac equation  in curved space for $m=0$  is conformal
invariant without any additional modifications~\cite{GMM,BD}.

    Furthermore, let us consider an $N$-dimensional homogeneous isotropic
space-time, choosing the metric in the form
    \begin{equation}
ds^2=g_{ik}dx^i dx^k = a^2(\eta)\,(d{\eta}^2 - d l^2) \,,
\label{gik}
\end{equation}
    where $d l^2=\gamma_{\alpha \beta} d x^\alpha d x^\beta $ is the metric
of an $(N-1)$-dimensional space of constant curvature $K=0, \pm 1 $.

    The complete set of solutions to Eq.~(\ref{Eqm}) in the metric~(\ref{gik})
may be found in the form
    \begin{equation}
\varphi(x) = \frac{\tilde{\varphi}(x)}{a^{(N-2)/2} (\eta)} \,
= a^{-(N-2)/2} (\eta)\, g_\lambda (\eta) \Phi_J ({\bf x})\,,
\label{fgf}
\end{equation}
    where
    \begin{equation}
g_\lambda''(\eta)+\Omega^2(\eta)\,g_\lambda(\eta)=0 \,,
\label{gdd}
\end{equation}
       \begin{equation}
\Omega^2(\eta)=(m^2 + (\xi - \xi_c) R) a^2 +\lambda^2 ,
\label{Ome}
\end{equation}
     \begin{equation}
\Delta_{N-1}\,\Phi_J ({\bf x}) = - \biggl( \lambda^2 -
\biggl( \frac{N-2}{2} \biggr)^2 K \biggr) \Phi_J  ({\bf x})\,,
\label{DFlF}
\end{equation}
    the prime denotes a derivative with respect to the conformal time $\eta$,
and $J$ is the set of indices (quantum numbers) numbering the eigenfunctions
of the Laplace-Beltrami operator $\Delta_{N-1}$
in ($N\!-\! 1$)-dimensional space.

    To perform quantization, let us expand the field $ \varphi(x) $
in the complete set of solutions~(\ref{fgf})
    \begin{equation}
\varphi(x)=\int \! d\mu(J)\,\biggl[ \varphi^{(+)}_J \,a^{(+)}_J +
\varphi^{(-)}_J \, a^{(-)}_J \,\biggr],
\label{fff}
\end{equation}
    where $d\mu(J)$ is a measure on the set of quantum numbers,
    \begin{equation}
\varphi^{(+)}_J (x)\!=\!\frac{g_\lambda(\eta)\,\Phi^*_J({\bf x})}
{\sqrt{2}\, a^{(N-2)/2}(\eta)}, \ \ \varphi^{(-)}_J(x) =
\bigl(\varphi_J^{(+)}(x)\! \bigr)^* \!,
\label{fpm}
\end{equation}
    and require that the standard commutation relations hold for
$a^{(\pm)}_J\!$ and $\stackrel{*}{a}\!{\!}^{(\pm)}_J$.

    A corpuscular interpretation of a quantized field based on the method
of Hamiltonian diagonalization by Bogoliubov's transformations have been
proposed for the gravitational external field in paper~\cite{GribMamayev69}.

    We give the following definition of particles in the external
field~\cite{GMM}, p.~46:
``In the framework of this interpretation one calls as particles
(quasiparticle) creation-annihilation operators at moment $t$ those operators
in terms of which the Hamiltonian of a quantized field is diagonal at the
moment~$t$.
    Therefore a quasiparticle is interpreted as an energy quantum and the
measurement of the number of quasiparticles is connected with the measurement
of energy.
    Indeed, the measurement theory in quantum mechanics requires that in
the result of measuring of some physical quantity a system would found itself
in a proper state of the corresponding operator.
    Therefore the measurement of energy inevitably transfers the system in
a proper state of the Hamiltonian.
    To find this state the Hamiltonian must be diagonalized.''

    Let us build the Hamiltonian as the canonical one for the variables
$\tilde{\varphi}(x)$ and $\tilde{\varphi}^*(x)$, for which the equation
of motion does not contain their first-order derivatives with respect to
the time $\eta$.
    Recall that the equations of motion do not change after adding a full
divergence $\partial J^i / \partial x^i$ to the Lagrangian density~$L(x)$.
    Let us choose, in the coordinate system $(\eta, {\bf x})$, the vector
    \begin{equation}
(J^i)=(\,\sqrt{\gamma} c \tilde{\varphi}{}^* \tilde{\varphi} (N-2)/2, 0,
\ldots , 0\,),
\label{JIdef}
\end{equation}
    where $\gamma={\rm det}(\gamma_{\alpha \beta})$, $c=a'/a$.
    Then, using the Lagrangian density
$ L^{\Delta}(x)=L(x) + \partial J^i / \partial x^i $,
we obtain for the momenta canonically conjugate to $\tilde{\varphi}$ and
$\tilde{\varphi}^*$:
    \begin{equation}
\pi \equiv \frac{\partial L^{\Delta}}{\partial \tilde{\varphi}'}=
\sqrt{\gamma}\, \tilde{\varphi}^*{}' \ , \ \ \
\pi_* \equiv \frac{\partial L^{\Delta}}{\partial
\tilde{\varphi}^{* \prime}} = \sqrt{\gamma}\, \tilde{\varphi}',
\label{imp}
\end{equation}
    respectively.
    Integrating the Hamiltonian density $h(x)=\tilde{\varphi}{}' \pi +
\tilde{\varphi}^{* \prime} \pi_* - L^{\Delta}(x) $
over the hypersurface $\Sigma$: $\eta = {\rm const} $,
we obtain the following expression for the canonical Hamiltonian:
    \begin{eqnarray}
H(\eta) \!=\!
\int_\Sigma d^{N-1}x \, \sqrt{\gamma} \, \biggl\{
\tilde{\varphi}^{* \prime} \tilde{\varphi}^\prime
+ \gamma^{\alpha \beta} \partial_\alpha\tilde{\varphi}^*
\partial_\beta \tilde{\varphi} {} +
\nonumber                 \\
\biggl[ (m^2 \! + \xi R ) a^2 - \frac{N\!-\!2}{4} \left(2c'+
(N \!-\!2)c^2\right) \biggr] \tilde{\varphi}^* \tilde{\varphi} \biggr\}.
\label{hx}
\end{eqnarray}

    The Hamiltonian~(\ref{hx}) may be written in terms of the operators
$ a_J^{(\pm)} $ and $ \stackrel{*}{a}\!{\!}_J^{(\pm)} $ in the following way:
    \begin{eqnarray}
H(\eta)\! &=& \! \int \! d\mu(J) \, \biggl[ E_J(\eta)
\left( \stackrel{*}{a}\!{\!}^{(+)}_J a^{(-)}_J +
\stackrel{*}{a}\!{\!}^{(-)}_{\bar{J}} a^{(+)}_{\bar{J}} \right) +
  \nonumber \\
&+& F_J(\eta)  \stackrel{*}{a}\!{\!}^{(+)}_J a^{(+)}_{\bar{J}} +
F^*_J(\eta) \stackrel{*}{a}\!{\!}^{(-)}_{\bar{J}} a^{(-)}_J  \biggr],
\label{H}
\end{eqnarray}
\vspace*{-2mm}
    where
\vspace{-1mm}
\begin{equation}
E_J=\frac{|g_\lambda'|^2+ \Omega^2 |g_\lambda|^2}{2},
\ \ \ F_J=\frac{\vartheta_{\!J}}{2} \bigl[ g_\lambda'{}^{\! 2} +
\Omega^2  g_\lambda^{\, 2} \bigr],
\label{EJFJ}
\end{equation}
and we have chosen such eigenfunctions $\Phi_{\!J}({\bf x})$ that,
for arbitrary $J$, there is such $\bar{J}$ that
$
\Phi_{\!J}^*({\bf x}) = \vartheta_{\!J} \Phi_{\!\bar{J}}({\bf x}),
\ |\vartheta_J|=1  \,,
$
($\ \bar{\!\!\bar{J}} = J$, \ $\vartheta_{\!\bar{J}}=\vartheta_{\!J}$).
    Such a choice is possible due to completeness and orthonormality of
the set $\Phi_{\!J}({\bf x})$.

    According to the Hamiltonian diagonalization method~\cite{GMM}
(the nonconformal case see in~\cite{Pavlov2001,Pavlov2002(IJMPA)}),
the functions $g_\lambda(\eta)$ should obey the following initial conditions:
    \begin{equation}
g_\lambda'(\eta_0)=i\, \Omega(\eta_0)\, g_\lambda(\eta_0) \,, \ \
\ |g_\lambda(\eta_0)|= \Omega^{-1/2}(\eta_0)\,.
\label{icg}
\end{equation}

    If the quantized scalar field is in the vacuum state at
the instant~$\eta_0$, then the number density of the pairs of particles
created up to the instant~$\eta $ can be evaluated (for $K=0$) as~\cite{GMM}
    \begin{equation}
n(\eta) = \frac{B_N}{2 a^{N-1}} \int \limits_0^\infty \! S_\lambda(\eta)\,
\lambda^{N-2}\, d \lambda,
\label{nN}
\end{equation}
where $B_N=\left[2^{N\!-3} \pi^{(N \!-1)/2} \Gamma((N\!-1)/2) \right]^{-1}\!,$
\ $\Gamma(z)$ is the gamma function, and
    \begin{equation}
S_\lambda(\eta) = \frac{\left| g'_\lambda (\eta ) -
i \Omega \, g_\lambda (\eta ) \right|^2}{4 \Omega } \,.
\label{Sgg}
\end{equation}
    As shown in~\cite{Pavlov2001}, $ S_\lambda \sim \lambda^{-6} $,
and the integral in~(\ref{nN}) converges for $N<7$.
    So the density of particles created in four dimensional Friedmann
space-time is finite.
    Note that the first use of the method of diagonalization of the
Hamiltonian in paper~\cite{Imamura60} led to the infinite density of created
quasiparticles.
    However there not the canonical Hamiltonian for
fields~$\tilde{\varphi}(x)$, but the Hamiltonian which is obtained from
the metrical energy momentum tensor for the scalar field with minimal coupling
was used.
    In the work~\cite{Parker69} created particle were defined for the so called
adiabatic vacuum state in such a way that the corresponding Hamiltonian was
not diagonal at any moment of time.
    This prevented to obtain finite results for particle creation in Friedmann
space-time.

    Note that the equation~(\ref{gdd}) by use of the transformation
$ g(\eta) = \exp z(\eta) $ is going to  the following equation for the function
$v(\eta) \equiv z'(\eta) $
    \begin{equation}
v'(\eta)+ v^2(\eta) + \Omega^2(\eta)=0 ,
\label{Riccati}
\end{equation}
    being the Riccati equation of the general type the solutions of which
cannot be expressed in finite form in elementary functions.
    So the number of scale factors~$a(\eta)$ leading to exact solutions is
relatively small.
    In cases when exact solution can be found it usually is expressed through
special functions --- hypergeometric functions, Bessel functions etc.

\section{\large Particle creation in cosmological models
with $ p = w \varepsilon $}
\hspace{\parindent}
    The Einstein equations
    \begin{equation}
R_{ik} - \frac{1}{2} g_{ik} R = - 8 \pi G T_{ik} ,
\label{Ein}
\end{equation}
    with the energy-momentum tensor of the background matter
$ T^i_k = {\rm diag}\, ( \varepsilon, -p, \ldots , -p\,) $
in metric~(\ref{gik}) have the form
    \begin{eqnarray}
\frac{c^2 +K}{a^2} = \frac{16 \pi G\, \varepsilon }{(N-1)(N-2)} \,,
 \ \ \ \ \nonumber \\
- \frac{1}{a^2} \Bigl[\, c^{\, \prime} + \frac{N-3}{2} \left( c^2 +K \right)
\Bigr] = \frac{8 \pi G p }{N-2}.
\label{Ein1}
\end{eqnarray}
    From~(\ref{Ein1}) it follows that for $ p =w \varepsilon $ where
$ w = {\rm const } $,
the energy density of the background matter is changing according to the law
$\varepsilon \sim  a^{- (1+w)(N-1)}$,
i.e. is decreasing with the increasing of $a$ and if $w>-1$ it is constant,
for $w=-1$ it increases with increasing $a$ for $w<-1$.

    For $K=0$ and $w > -(N\!-3)/(N\!-1)$ from~(\ref{Ein1}) one obtains
    \begin{eqnarray}
a = a_0\, t^{q} = a_1 \eta^{\beta}, \ \ \ q = \frac{2}{(N \!-1)(w+1)} \,,
\nonumber \\
\beta = \frac{q}{1 - q} , \ \ \ q \in (0, 1).
\label{ate}
\end{eqnarray}
    Expression for the number of pairs $N_a(t)= n(t)a^3(t) $ created in
the volume $a^3(t)$ up to the moment of time~$t$ can be written as
    \begin{equation}
N_a (t) = b^{(0)}_{\,q}(t) \! \cdot \Bigl( \frac{a(t_C)}{t_C}
\Bigr)^{\! N-1} \!,
\label{Nat}
\end{equation}
    where $t_C=1/m$  is the Compton time.
    Then $ b^{(0)}_{\,q}(t) / (1-q)^{N-1} $ is the coefficient of
proportionality for the number of created particles and the number of causally
disconnected regions $ N_c(t) = ((1\!-q)\, a(t)/ t)^{N-1} $ at the Compton time
after the Big Bang.

\begin{figure}[h]
 \resizebox{8cm}{!}{\includegraphics{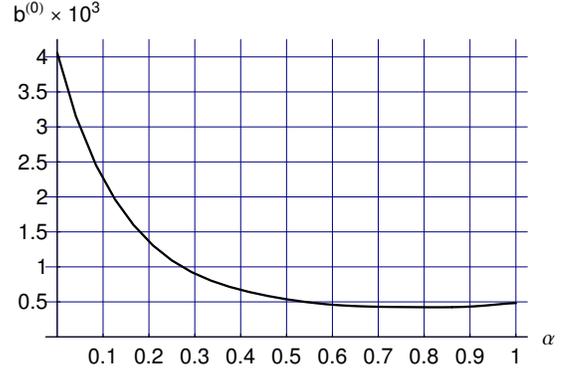}}
\caption{$b^{(0)}$ for conformal scalar particles and $a(t) \sim t^\alpha$.}
  \label{grBq}
\end{figure}

    The results of calculations~\cite{GribPavlov2008} of the number of created
particles for the scalar particles with conformal coupling don't depend on
the moment of time when one puts the initial condition as well as on time of
observation of created particles if these times are much smaller and later
than the Compton one.
    On Fig.~\ref{grBq} the results of numerical calculations of the
coefficient $b^{(0)}$ for $N=4$
    for different scale factors $a(t)=a_0\, t^\alpha$ are given.
    From~(\ref{Nat}) one can see the effect of connection of the number of
created particles with the number of causally disconnected parts on the
Friedmann Universe at the Compton time of its evolution.

     The results for nonconformal scalar field can radically depend on
the initial moment of time (see Eqs.~(\ref{gdd}), (\ref{Ome})).
    However for the condition $ (\xi_c - \xi) R > 0 $, for the case of
the minimally coupled scalar field in the dust Universe
$p=0$\ ($a(t) \sim t^{2/3}$), the initial moment can be defined by the
condition: $\Omega^2 \ge 0$ for any $\lambda$.
    In the other case (negative square of energy) one can obtain change of
the vacuum taking into account self interaction as it occurs for spontaneous
breaking of symmetry.

\begin{figure}[h]
 \resizebox{8cm}{!}{\includegraphics{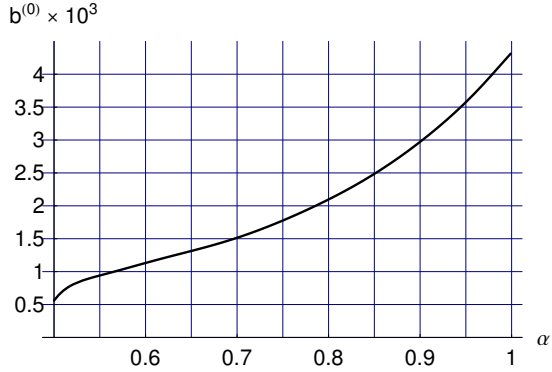}}
\caption{$b^{(0)}$ for scalar particles with minimal coupling and
$a(t)\sim t^\alpha$.} \label{grNBq}
\end{figure}

    Here we take for nonconformal scalar field initial value of $t_0$
so that $ m^2 + (\xi - \xi_c) R(t_0)=0 $.
    The results of numerical calculations~\cite{GribPavlov2008} of
the coefficient $b^{(0)}$ for the scalar field with minimal coupling and
the scale factor $a(t)=a_0\, t^\alpha$ are given on Fig.~\ref{grNBq}.

    In the interval $ q \in (0, 1)$ the following two cases has the exact
solutions of equations for the scalar field.

\subsection{The exact solution for $ a_0 \sqrt{t} = a_1 \eta $}

    The  Friedmann model with such scale factor is very important for
applications because for $K=0$ in four-dimensional space-time
it corresponds to the radiationally dominated Universe.
    The initial conditions for the conformally coupled scalar field
can be put at $\eta \to 0$ i.e. close to the singularity $a=0$
    \begin{equation}
|g_\lambda(0)|= 1/\sqrt{\lambda} \,, \ \ \ \ \
g_\lambda^{\prime}(0)= i \lambda g_\lambda(0).
\label{igk}
\end{equation}
    The solution of equation~(\ref{gdd}) with conditions~(\ref{igk})
can be written in the form
  \begin{eqnarray}
g_\lambda(t) = \frac{  e^{i (mt + \alpha_0) } }{ \sqrt{\lambda} }
\biggl[ \Phi \Bigl( \frac{1}{4} - \frac{i}{2} \delta^2 ,
\frac{1}{2} ; - i 2 m t \Bigr) + \nonumber \\
i 2 \delta \sqrt{m t} \ \Phi \Bigl( \frac{3}{4} -
\frac{i}{2} \delta^2 , \frac{3}{2} ; - i 2 m t \Bigr) \biggr],
\label{so1}
\end{eqnarray}
    where $ \Phi ( a ,\, b ;\, z ) $ is the degenerate hypergeometric
Kummer function,
$\delta \equiv \lambda /(m a(t_C)) $ has the sense of the physical moment at
the Compton time $t_C=1/m $ in units $m$.\,
$\alpha_0 $ is arbitrary real constant.
    The representation for~(\ref{so1}) through functions of the parabolic
cylinder is given at~\cite{MMSt}.
   The asymptotic value $b^{(0)}_{1/2} \approx 5,3 \cdot 10^{-4} $~\cite{MMSt}.

\subsection{The exact solution for $a_0 t^{1/3}\!=\!a_1 \sqrt{\eta}$}

    The model with such scale factor for $K=0$ and $N=4$ is the Universe
with limiting rigid $p=\varepsilon$ equation of state.
    The solution of the equation~(\ref{gdd}) with $\xi=\xi_c$ and initial
conditions~(\ref{igk}) for such scale factor~\cite{MMSt}:
    \begin{eqnarray}
g_\lambda(t) = \frac{\pi \delta^2 e^{i\alpha_0}}{\sqrt{3 \lambda}}\,
\sqrt{ (m t)^{2/3} + \delta^2} \times \nonumber \\
\left[\, C_1(\delta) J_{1/3} \left( \bigl( (m t)^{2/3} +
\delta^2 \bigr)^{3/2} \right) \right. + \nonumber \\
+\left. C_2(\delta)\, J_{-1/3} \Bigl( \bigl( (m t)^{2/3} +
\delta^2 \bigr)^{3/2} \Bigr) \right],
\label{so2}
\end{eqnarray}
    where $J_\nu(x)$ --- Bessel functions
    \begin{eqnarray}
C_1(\delta) = J_{2/3} \! \left( \delta^3 \right) +
i J_{-1/3}\! \left( \delta^3 \right),
\nonumber \\
C_2(\delta) = J_{-2/3}\! \left( \delta^3 \right) -
i J_{1/3}\! \left( \delta^3 \right)\!, \ \
\delta \equiv \frac{\lambda}{m a(t_C)} .
\label{C12d}
\end{eqnarray}
    The asymptotic value for $b^{(0)}_{1/3} \approx 8,1 \cdot 10^{-4}$.

\subsection{The exact solution for $a(t) \sim t $}

    For $ w = - (N-3)/(N-1) $ one obtains from the Einstein
equations~(\ref{Ein1}) that $ a=a_0 t = a_1 e^{a_0 \eta}$.
    If $a_0=1$ and $K=-1$, then $\varepsilon =0 $ and the metric~(\ref{gik})
with such scale factor is flat while coordinates $x^k$ describe the part of
the Minkowski space.
    The four dimensional hyperbolic space-time with $a(t)=t$ is known as
the Milne Universe.

    In metric(\ref{gik}) with $ a=a_0 t $ the solution of equation~(\ref{gdd})
with $\xi=\xi_c$ and initial conditions~(\ref{icg}) put at
the moment $mt \to 0$ has the form
    \begin{equation}
g_\lambda(t) = \frac{\sqrt{\lambda}}{a_0}\,
\Gamma \Bigl( \frac{i \lambda}{a_0} \Bigr)
J_{i \lambda / a_0}\!(mt)\, e^{i \alpha_0},
\label{gat}
\end{equation}
    where $\Gamma(z)$ is the gamma function,
$\alpha_0 $ is the arbitrary real constant.
    Independently from $a_0$ and $K$ the asymptotic at $mt \gg 1$
value of the density of created quasiparticles for $N=4$ is equal to
$n(t) \approx m/(512 \pi t^2)$.
    So different from zero result takes place even for the Milne Universe
where gravitational field is absent and one cannot have creation of real
particles!
    Really the space-time analysis of particle creation using correlation
function of the created pair of particles introduced in~\cite{MT} shows that
the corresponding correlation function of the created pair is exponentially
small at distances larger than the Compton length of the particle.
    This means that created quasiparticles in this case are virtual pairs
with the characteristic correlation length~$1/m$.

\subsection{De Sitter space-time}

    This is a space of constant curvature.
    It is a solution of Einstein equations in empty space with nonzero
cosmological constant.
    Using the special choice of coordinates in De Sitter space one can write
its metric in the form~(\ref{gik}) with scale factor 1--3)
for the De Sitter space of the first type and 4)
for the De Sitter space of the second type:
    \begin{eqnarray}
1) \, a_1 \, e^{Ht} = \frac{-1}{H \eta}, \ \
2) \, \frac{\cosh Ht}{H}  = \frac{1}{H \sin \eta}, \ \
\nonumber \\
3) \, \frac{\sinh Ht}{H}  = \frac{-1}{H \sinh \eta}, \ \
4) \, \frac{\cos Ht}{H}  = \frac{1}{H \cosh \eta}.
\label{dSrazn}
\end{eqnarray}
    In all these cases the equation for the scalar field~(\ref{gdd})
has the exact solution.
    In case~1),
$t \in (-\infty,+\infty) \ \Leftrightarrow \ \eta \in (-\infty, \, 0)$,
the solution of equation~(\ref{gdd}) with conditions~(\ref{icg})
for $\eta_0 \to -\infty$ has the form
    \begin{eqnarray}
g_\lambda(\eta)= \sqrt{-\frac{\pi \eta}{2}}
\ e^{{\textstyle \frac{\pi}{2}}\, {\rm Im}\, \nu}
H_\nu^{(2)}(-\lambda \eta)\, e^{i \alpha_0}, \ \ \
\nonumber \\
\nu=\sqrt{\frac{1}{4} - \frac{m^2+(\xi-\xi_c)R }{H^2} },
\label{GR70dSg}
\end{eqnarray}
    where $ H_\nu^{(2)}(z) $ is the Hankel function,
$\alpha_0$ is arbitrary real constant.
    As it was shown in~\cite{MT} (and for nonconformal case
for $ m^2 \ge (\xi_c - \xi) R $ in~\cite{Pavlov08})
by the method of the space-time correlation function the created pairs
must be interpreted as virtual pairs.
    Absence of creation of real particles in De Sitter space is confirmed by
the local form of the vacuum energy-momentum tensor and zero imaginary part
of the effective Lagrangian~\cite{GMM}.

    In cases 2 -- 4 also for the same scale factors but taking
$\eta \to \gamma \eta, \ \gamma = {\rm const} $ the solution also can be
expressed through hypergeometric functions
(see \S\,9.5 in~\cite{GMM} and~\cite{Pavlov2013}).
    For $\gamma \ne 1$ these  scale factors don't describe the De Sitter space.
    For example the space-time with the scale factor
$ a(\eta) = 1 / (H \cosh \gamma \eta) = \sin (\gamma H t) / H $
has the evolution between two singularities.

\subsection{Particle creation in cosmology with phantom matter}

    For $w< -1$ the exact solution of equation~(\ref{gdd}) exists for
the value $w=-(N+1)/(N-1)$ \ ($ w = -5/3$, if $N=4$).
    The scale factor of metric in this case is
$ a = a_0 / (-t) = a_1 / \sqrt{-\eta} $.
    For $t \to -0$ there is a singularity of the Big Rip.
    The solution of equation~(\ref{gdd}) satisfying the conditions~(\ref{icg})
for $\eta_0 \to -\infty$ has the form
    \begin{eqnarray}
g_\lambda(\eta)= -2i \eta \sqrt{\lambda}\, \exp \Bigl(\!
-\frac{\pi m^2 a_1^2}{4 \lambda} + i ( \lambda \eta +\alpha_0)\! \Bigr)
\nonumber \\
\times \Psi \Bigl(\! 1+ \frac{i m^2 a_1^2}{2 \lambda}\, , 2\, ;
-2 i \lambda \eta \! \Bigr),
\label{solBR}
\end{eqnarray}
    where $ \Psi ( a ,\, b ;\, z ) $ is the degenerate hyperbolic Tricomi
function, $\alpha_0 $ is the arbitrary real constant.
    For the density of particles created in $N=4$ for $t \to -0 $,
one obtains (look~\cite{Pavlov09}) $ n = m^3/24 \pi^2$.
    In spite of the fact that the general number of created particles
$N_a(t) = n(t) a^{3} (t) $ in Lagrange volume $ a^{3}(t)$
is unboundedly increasing for $t \to -0$,
the back reaction of particles creation and vacuum polarization of a massive,
conformally coupled scalar field on the space-time metric can be
neglected in the whole region where one can apply the approach of quantum
field theory in curved space-time~\cite{Pavlov09}.

\section{\large Exact solutions in cosmological models
with $p/\varepsilon \ne {\rm const}$}
\hspace{\parindent}
    Let us make a short review of scale factors leading to exact solutions of
the equation~(\ref{gdd}).
    Solutions for metrics
    \begin{equation}
a(\eta) = A + B \tanh \gamma \eta, \ \
a(\eta) = \sqrt{ A + B \tanh \gamma \eta } ,
\label{TReshg1}
\end{equation}
$A, \ B, \ \gamma = {\rm const}$
    are expressed through the hypergeometric function
(see~\cite{GMM}, \S\,9.5, \cite{BD}, \S\,3.4).

    Metric with the scale factor $ a(\eta) = \sqrt{ a \eta^2 + b \eta} $
in four dimensional space-time for $\eta \ll b/a$ corresponds to the limiting
rigid equation of state $p=\varepsilon$.
    For $\eta \gg b/a$ ($K=0$) this scale factor corresponds to the
radiation dominated Universe $p=\varepsilon/3$.
    The solution of equation~(\ref{gdd}) with $\xi=\xi_c$
for this scale factor can be expressed through the degenerate
hypergeometric Kummer function.

    Homogeneous isotropic space-time with the scale factor
    \begin{eqnarray}
a(\eta) = a_1 \tan \gamma \eta =
a_1 \sqrt{ \exp\left( \frac{2 \gamma t}{a_1} \right) -1 } ,
\nonumber \\
\eta \in \left( 0, \, \frac{\pi}{2 \gamma} \right)\!, \ \ \
t \in (0, \, +\infty )
\label{TReshgtg}
\end{eqnarray}
    is describing the the radiation dominated Universe
$ a(t) \approx \sqrt{2 \gamma a_1 t} $ for small times,
for larger times ---  exponentially expanding Universe
$ a(t) \approx a_1\, \exp \gamma t /a_1 $.
    The solution of equation~(\ref{gdd}) for the scalar field with conformal
coupling and initial values at $\eta \to 0$ is expressed through
the geometric function $ F ( a ,\, b ,\, c ;\, z ) $:
    \begin{eqnarray}
g(\eta) = \frac{e^{i \alpha_0}}{\sqrt{\lambda}}
(\cos \gamma \eta )^{ \left( 1+ \sqrt{ 1 - 4 m^2 a_1^2/\gamma^2}
\, \right) / 2 }
\nonumber \\
\biggl[ F \Bigl( \alpha , \beta , \frac{1}{2} ;\, \sin^2 \gamma \eta \Bigr) +
i \frac{\lambda}{\gamma} \sin \gamma \eta \times
\nonumber \\
\times F \Bigl(\! \alpha \!+\! \frac{1}{2},
\beta \!+\! \frac{1}{2}, \frac{3}{2} ; \sin^2 \gamma \eta \Bigr) \biggr],
\nonumber \\
\alpha, \beta  = \frac{1}{4}
\Biggl[ \!1 + \sqrt{1 \!- \frac{4 m^2 a_1^2}{\gamma^2}} \pm
\frac{2}{\gamma} \sqrt{ \lambda^2 \!- m^2 a_1^2} \Biggr].
\label{TReshgtg10}
\end{eqnarray}

\begin{figure}[h]
\centering
\includegraphics[width=57mm]{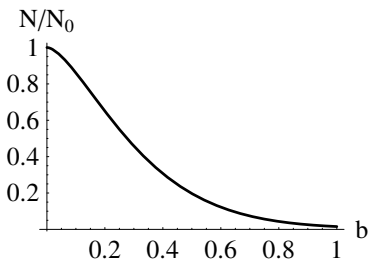}\\[-7pt]
\caption{The relation of number of the created particles in
model~(\ref{TReshg6}) with parameter~$b$ to the radiation dominated
case~($b=0$).}
\label{PCrRadb}
\end{figure}

    Note that if one finds the general solution~(\ref{gdd}) for the scalar
conformally coupled to the metric with the scale factor~$a(\eta)$
one can obtain by redefinition~$\lambda$ the general solution for
the scale factor $ \tilde{a} = \sqrt{ a^2(\eta) + b^2} $,
where $b={\rm const}$.
    For example take the model with the scale factor
    \begin{equation}
a(\eta) = \sqrt{ a_1^2 \eta^2 + b^2 } \,, \ \ \ \
-\infty < \eta < \infty,
\label{TReshg6}
\end{equation}
    which in asymptotic regions $ \eta \to \pm \infty $
corresponds (for $N=4$) to radiation dominated cosmology.
    The space is contracting to the minimal scale factor
$a(0)=b$, "is reflecting" and again is expanding.
    The solution of the equation~(\ref{gdd}) with initial
conditions~(\ref{icg}) defined by the condition of the diagonalization of
the Hamiltonian at the moment~$\eta=0$ like~(\ref{so1}) has the form
    \begin{eqnarray}
g_\lambda( \eta ) = \frac{ \exp i \left( \alpha_0 + ma_1 \eta^2/ 2 \right) }
{ \left( \lambda^2 + m^2 b^2 \right)^{1/4} } \times
\nonumber \\
\times \left[ \Phi \Biggl( \frac{1}{4} - \frac{i}{2}\, \delta^2 ,\,
\frac{1}{2}\, ; - i m a_1 \eta^2 \Biggr) \right. +
\nonumber \\[4pt]
+ \left. i  \eta \sqrt{\lambda^2 + m^2 b^2} \ \Phi \Biggl( \frac{3}{4} -
\frac{i}{2}\, \delta^2 ,\, \frac{3}{2}\, ; - i m a_1 \eta^2 \Biggr) \right]\!,
\label{TReshRad3}
\end{eqnarray}
$\delta^2 = \frac{\lambda^2 + m^2 b^2}{2 m a_1}$.
    The influence of the parameter $b$ on the intensity of particle
creation for the case $m=1$, $a=1/2$  is shown on Fig.~\ref{PCrRadb}.
    It is seen that for $b \to 0$ the number of created particles is going
to the known result for $b=0$.
    This shows the possibility of putting the initial conditions~(\ref{icg})
at $b=0, \, \eta \to 0$,
i.e. at the singularity for the radiation dominated Universe.

    Note that for some scale factors the solutions for the scalar field with
nonconformal coupling can be obtained by the redefinition of the mass and
momentum.
    The problem of description of such scale factors was solved
in~\cite{Pavlov2013}.

\section{Superheavy particles in the early Universe}
\hspace{\parindent}
     The total number of massive particles created in
Friedmann radiation dominated Universe (scale factor $a(t)=a_0\, t^{1/2}$)
 inside the horizon is (see~(\ref{Nat}))
    \begin{equation}
N_a = n^{(s)}(t)\,a^3(t) = b^{(s)}\,m^{3/2}\,a_0^3 ,
\label{NbM}
\end{equation}
    where $b^{(0)} \approx 5.3 \cdot 10^{-4}$ for scalar
and  $b^{(1/2)} \approx 3.9 \cdot 10^{-3}$ for spinor particles~\cite{GMM}.
    It occurs that $ N_a \sim 10^{80} $
for $ m \sim 10^{14} $\,GeV \cite{GMM,GD}.
    Calculations for creation from vacuum pairs of
superheavy particles with the mass of the Grand Unification scale in the
early radiation dominated Universe give the surprising result --- its number
is of the order of the observed Eddington-Dirac number.
    So putting the hypothesis of their decay on quarks and leptons in the early
Universe one obtains the observed number of protons and electrons.
    All this shows that one cannot neglect the effect of particle creation
in the early Universe!

    The radiation dominance in the end of inflation era is
important for our calculations.
    However this radiation is formed not by our visible particles.
    It is quintessence or some dark light particles not interacting with
ordinary particles.

    For the time ${t \gg m^{-1}} $ there is an era of going from the
radiation dominated model to the dust model of superheavy particles,
    \begin{equation}
t_X\approx \left(\frac{3}{64 \pi \, b^{(s)}}\right)^2
\left(\frac{M_{Pl}}{m}\right)^4 \frac{1}{m}  \,,
\end{equation}
    where $M_{Pl} \approx 1,2\cdot 10^{19}$\,GeV is Planck mass.
    If $m \sim 10^{14} $\,GeV,
$\ t_X \sim 10^{-15} $\,s for scalar and
$\ t_X \sim 10^{-17} $\,s for spinor particles.

   Let us define $d $, the permitted part of long-living
$X$-particles, from the condition: on the moment of
recombination $t_{rec} $ in the observable Universe one has
$
d\,\varepsilon_X(t_{rec}) =\varepsilon_{crit}(t_{rec})  \,.
$
    It leads to
\begin{equation}
d=\frac{3}{64 \pi \, b^{(s)}}\left(\frac{M_{Pl}}{m}\right)^2
\frac{1}{\sqrt{m\,t_{rec}}}\, .
\label{d}
\end{equation}
     For $ m=10^{13} - 10^{14} $\,GeV one has
$d \approx 10^{-12} - 10^{-14} $ for scalar and
$d \approx 10^{-13} - 10^{-15} $ for spinor particles.
    So the lifetime of the main part or all $X$-particles must be smaller
or equal than $t_X$.

     We give in Ref.~\cite{GribPavlov2002(IJMPD),GribPavlov2002(IJMPA)}
the model which can give: \ {\bf (a)} short-living $X$-particles decay in time
   $\tau_q < t_X $ (more wishful is
   $\tau_q \sim t_C \approx 10^{-38} - 10^{-35} $\,s,
i.e., the Compton time for $X$-particles); \
{\bf (b)} long-living particles decay with $\tau_l > 1/m $.

     If $\tau_l$ is larger than the time of breaking of the Grand Unification
symmetry it can be that some quantum number can be conserved leading to
some effective time
 $\tau_l^{\rm eff} > t_U \approx 4.3 \cdot 10^{17}$\,s
\  ($t_U $ is the age of the Universe).
    The small $ d \sim 10^{-15} - 10^{-12} $ part of long-living
$X$-particles with $\tau_l > t_U $ forms the dark matter.

    For $ t_l^{\rm eff} \le 10^{27}$\,s one must get a strong anisotropy
of ultra high energy cosmic rays in the direction to the center of
the Galaxy.
    However, experiments with cosmic rays don't show such an anisotropy
and one must suppose $ t_l^{\rm eff} > 10^{27}$\,s.
    The special conditions existing in vicinity of black holes can lead to
the collisions of superheavy particles with high energy and to their
decay there on ordinary particles as it was in the early Universe.
    Our estimates~\cite{GribPavlov2007AGN} show that such processes near
the black holes of active galactic nuclei can explain the stream of
ultrahigh energy cosmic rays with energy $\ge 10^{19}$\,eV observed
in experiments.

    The observed entropy in this scenario originates due to transformation of
$X$-particles into light particles: quarks, antiquarks and some particle
similar to $\Lambda^0$ in $K^0$-meson theory, having the same quantum
number as $X$.
    Baryon charge is created close to the time $t_q $, which can be equal to
the Compton time of
$X$-particles $t_C \sim 10^{-38} - 10^{-35} $\,s \cite{GribPavlov2008}.

\section{Conclusion
\label{secConcl}}
\hspace{\parindent}
    Particle creation from vacuum by the gravitation of the expanding
Friedmann Universe can explain the observable number of visible particles
as well as dark matter.
    In our scenario it is dark matter consisting of superheavy particles
with mass of the Grand Unification which is created first, ordinary particles
are created not by gravitation but by the conversion of particles of dark
matter into them at very high energy when high symmetry of Grand Unification
is present.
    Breaking of this symmetry leads to conservation of some part of originally
created dark matter particles which are stable today.

    This hypothesis can be checked in processes with elementary particles in
ergosphere of rotating black holes where collisions of particles can lead
to very high energy in the centre of mass frame so the situation of the early
Universe can be reproduced.

    In our papers~\cite{GribPavlov2007AGN,GribPavlov2008KLGN}
    it was shown
    1) conversion of dark matter particles with large mass into
visible particles with relatively small mass can lead to observation of
the products on the Earth as ultra high energy cosmic ray particles
inspite of the large red shift.
    2) Estimates of the process in
taking into account observable dark matter density lead to observable
flaws of ultra high energy cosmic rays.

    There are still some unsolved problems.

    a) In anisotropic space-time only in some special
case~\cite{Pavlov2002(IJMPA)} one can obtain the finite results by
the Hamiltonian diagonalization.
    So this technical problem is still unsolved.

    b) Quantization of minimal coupled field in general case leads to two
types of modes: usual oscillations and unstable growing modes.
    So that one must introduce indefinite metric for their quantization.
    Correct quantization needs taking into account nonlinear
terms (see paper~\cite{GD10} for some special case) leading to redefinition
of vacuum as it is done in spontaneous symmetry breaking theories.

{\bf Acknowledgments.}
    This work was supported by the Russian Foundation for Basic Research,
grant No. 15-02-06818-a.


\end{document}